\documentclass[aps,prb,twocolumn,amsmath, showpacs,floatfix,reprint]{revtex4-2}
\usepackage{amsmath, nccmath}
\usepackage{amssymb}
\usepackage{bm}
\usepackage{braket}
\usepackage{natbib}
\usepackage{leftidx}
\usepackage{graphicx}    
\usepackage{verbatim}   
\usepackage{color}      
\usepackage{subfigure}  
\usepackage{hyperref}   
\usepackage{dcolumn}    
\usepackage{textcomp}
\usepackage{float}
\usepackage{threeparttable}
\usepackage{titlecaps}
\usepackage{multirow}
\usepackage{lipsum}
\hypersetup{
	colorlinks=true,
	citecolor=blue,
	filecolor=black,
	linkcolor=blue,
	urlcolor=cyan
}
\usepackage{mathastext}
\usepackage{mathptmx}
\usepackage{enumitem}
\DeclareGraphicsExtensions{.png,.pdf,.tif}
\usepackage{pict2e}
\usepackage[dvipsnames]{xcolor}
\usepackage[normalem]{ulem}

\begin{document}

\title{Tuning of topological properties in the strongly correlated antiferromagnet Mn$_3$Sn via Fe doping}

\author{Achintya Low}

\author{Susanta Ghosh}

\author{Susmita Changdar}

\author{Sayan Routh}

\author{Shubham Purwar}

\author{S. Thirupathaiah}%
\email{setti@bose.res.in}
\affiliation{%
 Department of Condensed Matter and Materials Physics, S. N. Bose National Centre for Basic Sciences, Kolkata, West Bengal-700106, India.
}%


\begin{abstract}
Magnetic topological materials, in which strong correlations between magnetic and electronic properties of matter, give rise to various exotic phenomena such as anomalous Hall effect (AHE), topological Hall effect (THE), and skyrmion lattice. Here, we report on the electronic, magnetic, and topological properties of Mn$_{3-\it{x}}$Fe$_{\it{x}}$Sn single crystals ($\it{x}$=0, 0.25, and 0.35). Low temperature magnetic properties have been significantly changed with Fe doping. Most importantly, we observe that large uniaxial magnetocrystalline anisotropy that is induced by the Fe doping in combination with competing magnetic interactions at low temperature produce nontrivial spin-texture, leading to  large topological Hall effect in the doped systems at low temperatures. Our studies further show that the topological properties of Mn$_{3-\it{x}}$Fe$_{\it{x}}$Sn are very sensitive to the Fe doping.
\end{abstract}

\maketitle


\section{Introduction}

Magnetic topological materials are the illustrations of an interplay between magnetic and electronic states of matter, providing an important stage for illuminating several exotic phenomena such as the anomalous Hall effect (AHE)~\citep{PhysRev.95.1154,sinitsyn2007semiclassical, PhysRevB.74.214201}, the topological Hall effect (THE)~\citep{PhysRevLett.102.186602,PhysRevLett.106.156603,PhysRevLett.118.027201}, the skyrmionic lattice~\citep{yu2010real,nagaosa2013topological,doi:10.1126/science.1166767}, and etc. On the other hand, the kagome lattice in which atoms are arranged in star-like formation anticipates a geometrical frustration,  leading to noncollinear antiferromagnetic (AFM) ordering~\citep{moessner2006geometrical,PhysRevLett.110.184102,10.21468/SciPostPhysCore.5.1.007,PhysRevB.102.035127,PhysRevLett.119.087202,nagamiya1982triangular}. So far, several kagome intermetallics have been explored to a great extent due to their potential magnetic topological properties. For instance, Co$_3$Sn$_2$S$_2$ is a magnetic Weyl semimetal showing giant AHE in addition to chiral anomaly~\citep{liu2018giant,morali2019fermi}, Mn$_3$Sn(Ge) are time-reversal symmetry broken Weyl semimetals, despite being antiferromagnets,  show large AHE induced by the nonzero $k$-space Berry curvature~\citep{nakatsuji2015large,kuroda2017evidence,nayak2016large,yang2017topological}, Fe$_3$Sn$_2$ which is  a kagome ferromagnet generates skyrmionic bubbles in addition to the giant AHE~\citep{hou2017observation,wang2016anomalous}, YMn$_6$Sn$_6$ is a rare earth based kagome system showing several competing magnetic orders and large THE~\citep{ghimire2020competing}, and Gd$_3$Ru$_4$Al$_{12}$ posses low temperature skyrmion lattice induced by the magnetic frustration~\citep{hirschberger2019skyrmion}.

Skyrmions, the vortex-like spin texture formation in the real space, are topologically protected and characterized by their topological charge called the winding number~\citep{fert2017magnetic}. The skymions pursuit futuristic technological applications in the high-density data storage devices~\cite{Luo2021, nii2015uniaxial}, fine current controlled devices~\citep{finocchio2016magnetic}, and information processing devices~\cite{Luo2018}. There exist several systems showing skyrmion lattice that is originated from different mechanisms. For example, in noncentrosymmetric magnetic systems such as MnSi~\citep{nagaosa2013topological}, FeGe~\citep{yu2011near}, and FeCoSi~\citep{yu2010real} the skyrmion lattice formation was understood by the Dzyaloshinskii-Moriya interaction (DMI) under the strong spin-orbit coupling~\cite{DZYALOSHINSKY1958241,PhysRev.120.91}.  In the centrosymmetric magnetic systems such as La$_{1-x}$Sr$_x$MnO$_3$~\citep{yu2014biskyrmion} and Fe$_3$Sn$_2$~\citep{hou2017observation} the competition between magnetic dipole interactions and uniaxial magnetocrystalline anisotropy stabilizes the skyrmion lattice~\cite{PhysRevB.93.184413,PhysRevB.103.064414,doi:10.1073/pnas.1118496109}. In addition, recent studies show the existence of skyrmions in rare-earth based  intermetallics Gd$_2$PdSi$_3$ and Gd$_3$Ru$_4$Al$_{12}$ due to the magnetic frustration~\citep{kurumaji2019skyrmion,hirschberger2019skyrmion}.


Mn$_3$Sn is a kagome itinerant antiferromagnet with a N\'eel temperature of 420 K, has hexagonal crystal structure with a space group of P6$_3$/mmc~\citep{nakatsuji2015large, tomiyoshi1987electrical}. Here, the Mn atoms form kagome network in the $ab$ plane of the crystal, showing  chiral 120$^\circ$ inverse triangular spin structure stabilized by the DM interaction~\citep{park2018magnetic,PhysRevLett.119.087202}.
Due to a slight distortion in kagome lattice and as well off-stoichiometry of the system, usually, weak-ferromagnetism is present in this system~\citep{nagamiya1982triangular,nakatsuji2015large}. Moreover, low-temperature magnetic structure depends on the elemental ratio of Mn and Sn, annealing temperature, and crystal growth techniques. Thus, some studies report helical spin structure in Mn$_3$Sn at low temperatures~\citep{cable1993neutron,duan2015magnetic} while the other studies realized spin-glass state below 50 K~\citep{ohmori1987spin,feng2006glassy}. At room temperature,  Mn$_3$Sn shows noncollinear antiferromagnetic ordering with 120$^0$ inverse triangular spin structure~\cite{park2018magnetic,PhysRevLett.119.087202,nagamiya1982triangular,nakatsuji2015large} leading to large anomalous Hall effect. Particularly, Mn$_3$Sn is of great research interest as the triangular-coplanar magnetic order reshapes into a spiral-noncoplanar magnetic ordering with a finite net magnetization along the $c$-axis at a critical spin-reorientation transition temperature (T$_{SR}\approx$ 260 K)~\cite{park2018magnetic,PhysRevLett.119.087202,nakatsuji2015large} which is not found in Mn$_3$Ge~\cite{Kiyohara2016}. As a result, the large AHE is suppressed below T$_{SR}$ in Mn$_3$Sn but not in Mn$_3$Ge~\cite{Chen2021}.

Theoretically, Mn$_3$Sn is not expected to show topological Hall effect (THE) as the stoichiometric Mn$_3$Sn does not possess chiral-spin texture. However, there are few reports claiming the observation of a small topological Hall resistivity due to the field induced chiral-spin texture in polycrystalline Mn$_3$Sn~\cite{PhysRevB.99.094430} at low temperature and due to domain wall formation in the single crystalline Mn$_3$Sn at room temperature~\cite{Li2018, Li2019}. On the other hand, Fe$_3$Sn sharing similar crystal structure of Mn$_3$Sn is a ferromagnetic metal with an easy axis of magnetization in the $ab$ plane that can be shifted to the $c$-axis with doping~\citep{Sales2014, vekilova2019tuning}. Motivated from the polymorphic magnetic properties of Mn$_3$Sn and flexibility of manipulating the easy axis in Fe$_3$Sn, we have substituted Fe atoms at Mn sites of Mn$_3$Sn to enhance the topological properties of Mn$_3$Sn as Fe substitution introduces ferromagnetism to the system.

In this work, single crystals of Mn$_{3-\it{x}}$Fe$_{\it{x}}$Sn ($\it{x}$=0, 0.25, and 0.35) were systematically studied for their electrical resistivity, magnetic, and topological properties. While Mn$_3$Sn is found to be metallic in nature up to room temperature with a spin-reorientation driven kink at 260 K, with Fe doping the system shows magnetism induced metal-insulator (MI) transition at 240 K for $\it{x}$=0.25 and 150 K for $\it{x}$=0.35. In addition to MI transition, $\it{x}$=0.35 system shows disorder induced resistivity upturn with a minima at T$_m$=50 K.   As for the magnetic properties, Mn$_3$Sn is found to show a sudden drop in magnetization at a spin-reorientation transition temperature of 260 K and spin-glass-like transition below 40 K.  On the other hand, with Fe doping ferromagnetic transition has been introduced alongside with enhanced magnetic anisotropy. Also, anisotropic anomalous Hall resistivity has been induced at low temperatures with Fe doping. Particularly,  the out-of-plane Hall resistivity ($\rho_{zx}$) increases with decreasing temperature for all the compositions from 300 K down to their respective magnetic transition temperatures where a sudden change in Hall resistivity is noticed.  Though not much change in out-of-plane Hall resistivity is noticed with Fe doping at 2 K, the in-plane Hall resistivity ($\rho_{xy}$) is gigantically enhanced from -0.25 $\mu\Omega$-cm to 48 $\mu\Omega$-cm in going from $\it{x}$=0 to $\it{x}$=0.35.  Along with the anomalous Hall resistivity, a large topological Hall resistivity also is observed for both Fe doped systems at 2 K.

\section{Experimental Details}

Single crystals of Mn$_{3-\it{x}}$Fe$_{\it{x}}$Sn ($\it{x}$=0, 0.25, and 0.35) were prepared by self flux method~\cite{doi:10.1080/13642819208215073,sung2018magnetic,yan2019room}. First, Mn (Alfa Aesar $99.995\%$),  Fe (Alfa Aesar, $99.99\%$), and Sn (Alfa Aesar $99.998\%$) powders were taken with a ratio of (7-$\it{x}$) : $\it{x}$ : 3, mixed thoroughly before inserting into a preheated quartz tube,  and sealed under partial Argon pressure.  The mixture was then heated up to 1000$^o$C, slowly cooled down to 900$^o$C, and then was air quenched to room temperature by taking out the ampoule from furnace. In this way, we obtained shiny hexagonal and rod shaped single crystals with a typical size of 1.5 mm $\times$ 1 mm $\times$ 1 mm. X-ray diffraction (XRD) measurements  were done on different surfaces of the single crystals using Rigaku SmartLab equipped with 9 kW Cu K$_\alpha$ X-ray source. Elemental compositions of the crystals were calculated to be Mn$_{2.97}$Sn$_{1.03}$, Mn$_{2.74}$Fe$_{0.26}$Sn, and Mn$_{2.64}$Fe$_{0.36}$Sn using energy dispersive X-ray spectroscopy (EDXS) technique. For convenience we denote the compositions Mn$_{2.97}$Sn$_{1.03}$, Mn$_{2.74}$Fe$_{0.26}$Sn, and Mn$_{2.64}$Fe$_{0.36}$Sn  by $\it{x}$=0,  $\it{x}$=0.25, and $\it{x}$=0.35, respectively,  wherever applicable.

\begin{figure}
\includegraphics[scale=0.40]{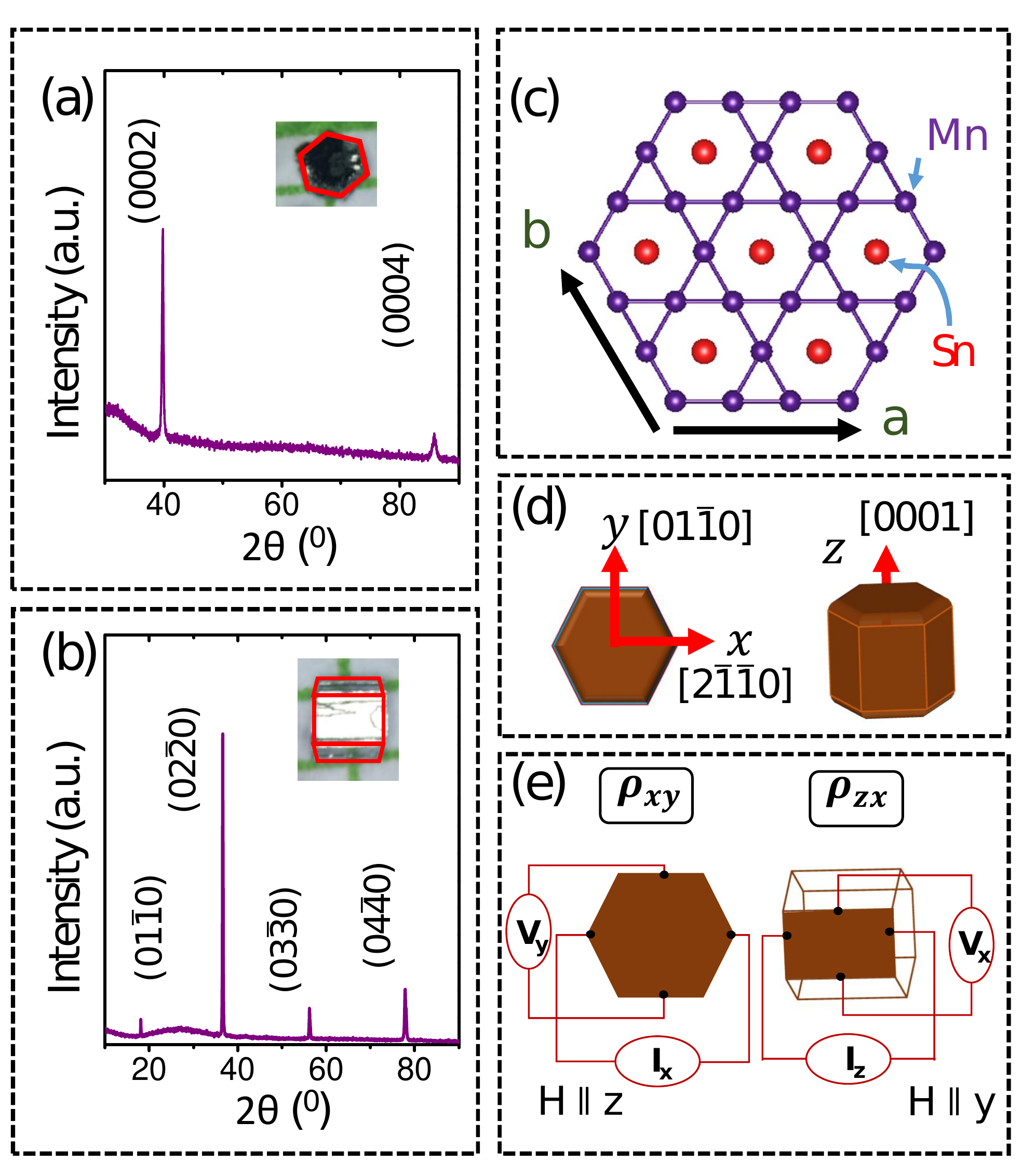}
\caption{\label{Fig1} (a) and (b) show typical XRD patterns of Mn$_3$Sn single crystal taken from two different orientations as shown in the inset images. (c) Schematic hexagonal lattice of $Mn_3Sn$ in the $ab$ plane where Mn atoms form kagome geometry and Sn atoms sit in the center of hexagon. In (d),  $x$, $y$, and $z$ -axes correspond to $[2\bar{11}0]$, $[01\bar{1}0]$, and $[0001]$ orientations, respectively.
Magnetotransport measuring geometry is shown in (e), where $\rho_{xy}$ is measured with current along the $x$-axis and external magnetic field applied along the $z$-axis to find Hall voltage along the $y$-axis and $\rho_{zx}$ is measured with current along the $z$-axis and field applied along the $y$-axis to find Hall voltage along the $x$-axis.}
\end{figure}

Electrical resistivity measurements were carried out in the linear four-probe method and Hall measurements were done in the measuring geometry shown in the schematic of Fig.~\ref{Fig1}(e). Magneto-transport and magnetization measurements were performed in Physical Properties Measurement System (9T PPMS, DynaCool, Quantum Design) using ETO and VSM options, respectively. Hall resistivity was measured for two different orientations as shown in Fig.~\ref{Fig1}(e). $\rho_{xy}$ stands for current along the $x$ direction and field along the $z$ direction where Hall voltage was measured along the $y$ direction. Similarly $\rho_{zx}$ stands for current along $z$ direction, field along the $y$ direction, and Hall voltage was measured along the $x$ direction.  To eliminate longitudinal voltage contribution due to any possible misalignment of the probes, the Hall resistivity was calculated by $\frac{\rho(H)-\rho(-H)}{2}$. Temperature dependent Hall resistivity measurements were done using both positive (H) and negative (-H) applied fields and then calculated using the formula $\frac{\rho(T, H)-\rho(T,-H)}{2}$.

\section{Results and Discussions}

\begin{figure}
\includegraphics[width=\linewidth]{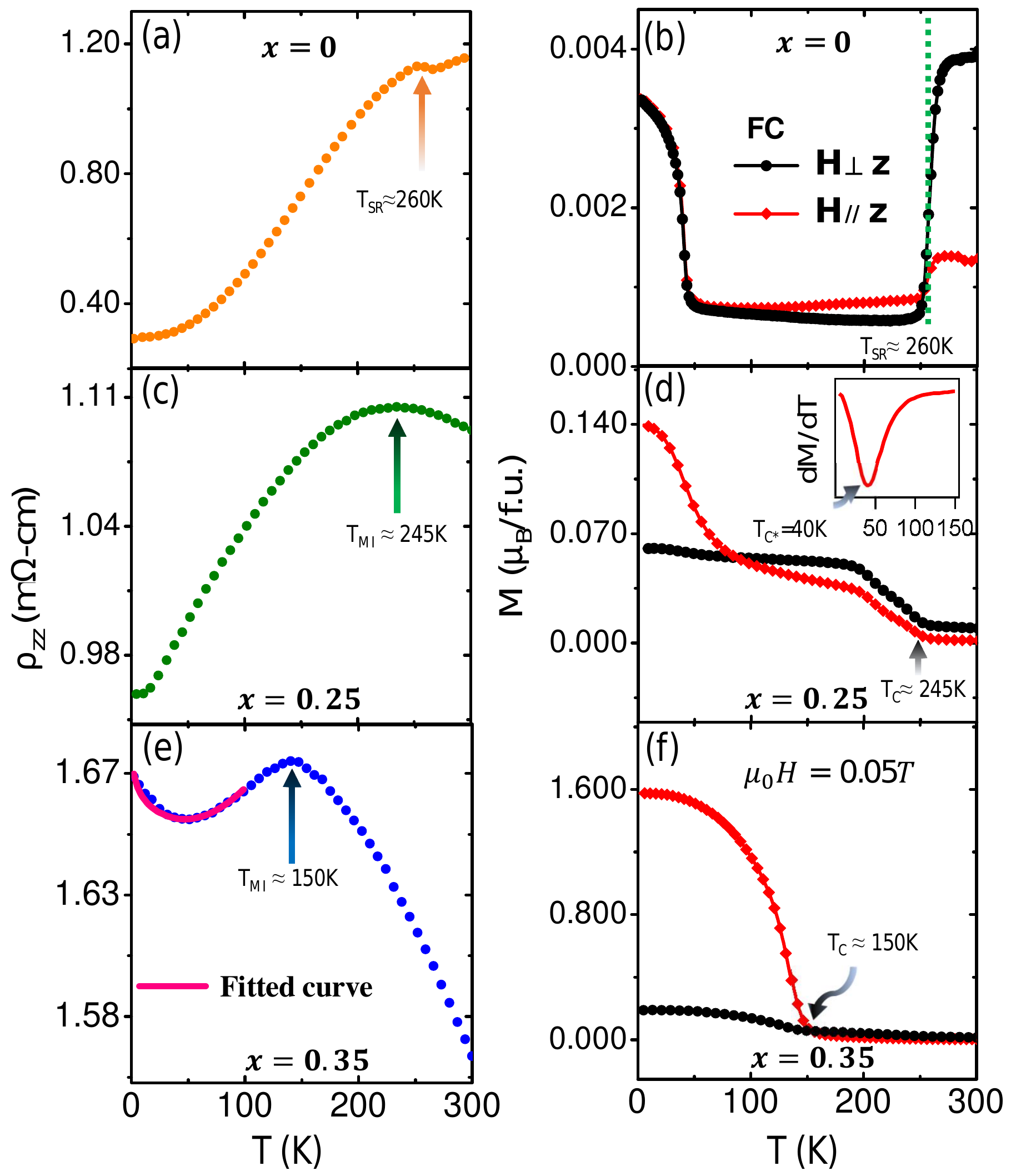}
\caption{\label{Fig2} Temperature dependent resistivity ($\rho_{zz}$) is measured with current along the $z$-axis for Mn$_{3-\it{x}}$Fe$_x$Sn where (a) $\it{x}$=0, (c)  $\it{x}$=0.25, and (e) $\it{x}$=0.35. (b), (d), (f) show temperature dependent magnetization for $\it{x}$=0, $\it{x}$=0.25, and $\it{x}$=0.35, respectively, measured in the field-cooled (FC) mode. Zero-field-cooled (ZFC) data is shown in Fig.S6 of the Supplemental Material~\cite{Supple}.}
\end{figure}

XRD patterns shown in Figs.~\ref{Fig1}(a) and ~\ref{Fig1}(b) represent intensity reflections of $(0002)$ and $(01\bar{1}0)$ hexagonal planes, respectively, taken from Mn$_3$Sn single crystal. Figs.~\ref{Fig2}(a), (c), and (e) depict zero-field out-of-plane resistivity ($\rho_{zz}$) measured between 2 and 300 K from Mn$_3$Sn, Mn$_{2.75}$Fe$_{0.25}$Sn, and Mn$_{2.65}$Fe$_{0.35}$Sn single crystals, respectively.  Overall, a metallic nature of resistivity is observed from Mn$_3$Sn, except that a kink has been noticed at $T_{SR}=260K$ where a spin-reorientation (SR) transition from in-plane noncollinear AFM order to an out-of-plane spin-spiral structure occurs~\cite{duan2015magnetic,sung2018magnetic}. Observation of kink in the electrical resistivity of Mn$_3$Sn is consistent with earlier reports~\cite{yan2019room,sung2018magnetic,tomiyoshi1987electrical}. On the other hand, we observe a metal-insulator (MI) transition in Mn$_{2.75}$Fe$_{0.25}$Sn at 240 K, which further reduced to 150 K in Mn$_{2.65}$Fe$_{0.35}$Sn as shown in Figs.~\ref{Fig2}(c) and ~\ref{Fig2}(e) with increasing Fe concentration.  Earlier, the authors found a MI transition at 265 K in the case of Mn$_{2.8}$Fe$_{0.2}$Sn~\cite{Low2021}.  In addition to the MI transition, in Mn$_{2.65}$Fe$_{0.35}$Sn, we identify low-temperature resistivity upturn with a minima at T$_m$=50 K. The resistivity upturn at low temperature could be due to (i) weak localization~\citep{PhysRevB.22.5142,dhara2016observation, PhysRevB.102.140409},  (ii) Kondo effect~\citep{kondo1964resistance,dhara2016observation}, or (iii) elastic electron-electron ($e-e$) scattering~\citep{lee1985disordered,dhara2016observation}.  In order to elucidate the mechanism of resistivity upturn,  we fit the data with relevant formulae involved in the above mentioned three mechanisms. We got the best fitting for the $e-e$ scattering which takes the form $\rho(T)=\rho_0-\alpha T^{\frac{1}{2}}+\beta T^2$,  as Fe doping induces disorder in addition to the electron carrier population. See the Supplemental Material~\cite{Supple} (see, also, references~\cite{nagamiya1982triangular, cable1993neutron, duan2015magnetic, PhysRevB.22.5142, kondo1964resistance, lee1985disordered, dhara2016observation, Lu2014, Xu2006} therein) for more details.

\begin{figure}
\includegraphics[width=\linewidth]{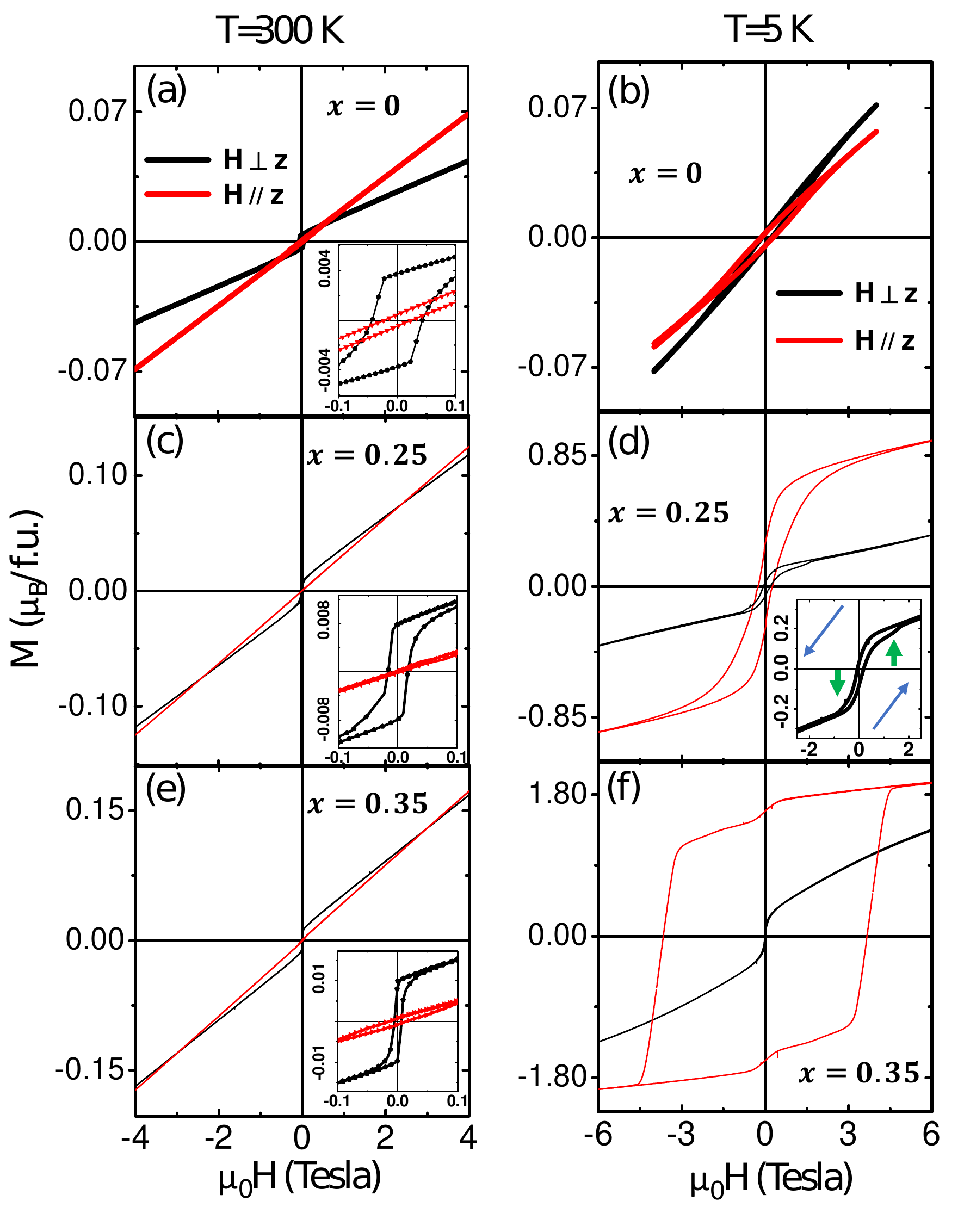}
\caption{\label{Fig3} (a), (c), and (e) show M(H) isotherms of Mn$_{3-\it{x}}$Fe$_x$Sn measured at 300 K for $\it{x}$=0, $\it{x}$=0.25, and $\it{x}$=0.35, respectively. (b), (d), and (f) show M(H) isotherms measured at 5 K for $\it{x}$=0, $\it{x}$=0.25, and $\it{x}$=0.35 respectively. Zoomed-in images of respective isotherms are show in the insets. Inset of (d) points to the asymmetric hysteresis (see the text for more details).}
\end{figure}

To witness the effect of Fe doping on the magnetic properties, we performed temperature dependent magnetization M(T) with field applied parallel ($H\parallel z$) and perpendicular ($H\perp z$) to the $z$-axis. From  the M(T) data of Mn$_3$Sn as shown in Fig.~\ref{Fig2} (b) we find a sudden drop in the in-plane magnetic moment at T$_{SR}$=260 K due to spin transformation from inverse triangular to helical or spiral configuration~\citep{yan2019room,sung2018magnetic}. Similarly, we find drop in the out-of-plane magnetic moment as well at T$_{SR}$=260 K but not as large as the in-plane. Reducing the sample temperature to below 40 K, we observe increasing in-plane and out-of-plane magnetic moments due to spin-glass-like transition~\citep{feng2006glassy,sung2018magnetic}. With Fe doping, in the case of $\it{x}$=0.25,  isotropic increase in magnetization is observed [see Fig.~\ref{Fig2}(d)] for both $H\parallel z$ and $H\perp z$ field orientations with a ferromagnetic-like transition at a T$_{C}$=240 K. With further lowering sample temperature the in-plane magnetic moment gets saturated, while the out-of-plane magnetization show additional ferromagnetic-like transition at T$_{C^*}$=40 K, enhancing the anisotropy of the magnetic structure at low temperatures.  Further, with more Fe doping, i.e., from  $\it{x}$=0.35 we see a ferromagnetic-like transition at $T_C$=150 K for both orientations though the out-of-plane magnetization (1.6$\mu_B/f.u.$) is much stronger compared to in-plane magnetization (0.3$\mu_B/f.u.$) at 2 K. Decreasing T$_C$ with increasing Fe concentration is consistent with a previous report~\cite{Felez2017}. In conjunction,  from the magnetization data shown in Figs.~\ref{Fig2}(b), (d), and (f) and the electrical resistivity shown in Figs.~\ref{Fig2}(a), (c), and (e) we find that the metal-insulator transition occur nearly at the same temperature of ferromagnetic transition in Mn$_{2.75}$Fe$_{0.25}$Sn and  Mn$_{2.65}$Fe$_{0.35}$Sn,  suggesting that ferromagnetism could be a plausible origin of MI transition in these systems.

\begin{figure}
\includegraphics[width=\linewidth]{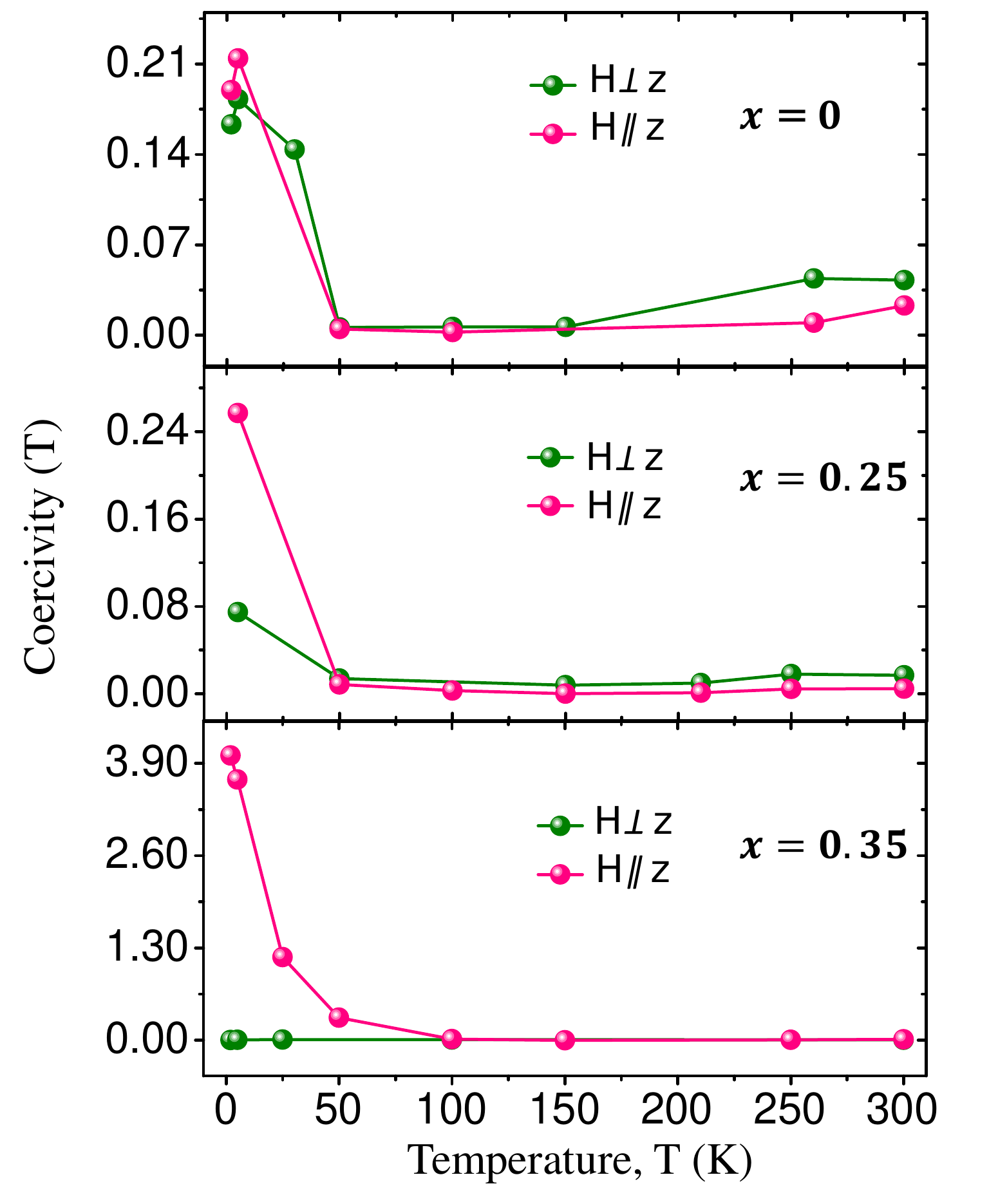}
\caption{\label{Fig4} Coercivity plotted as a function of temperature for both H $\parallel z$ and H $\perp z$. Top panel shows the data from Mn$_3$Sn, middle panel shows the data from Mn$_{2.75}$Fe$_{0.25}$Sn, and the bottom panel shows the data from Mn$_{2.65}$Fe$_{0.35}$Sn.}
\end{figure}

Fig.~\ref{Fig3} depicts magnetization isotherms M(H) measured at various sample temperatures for both $H\parallel z$ and $H\perp z$ orientations. Figs.~\ref{Fig3}(a), ~\ref{Fig3}(c), and ~\ref{Fig3}(e) show isotherms of $\it{x}$=0, $\it{x}$=0.25, and $\it{x}$=0.35 samples, respectively, measured at 300 K and Figs.~\ref{Fig3}(b), ~\ref{Fig3}(d), and ~\ref{Fig3}(f) show isotherms of $\it{x}$=0, $\it{x}$=0.25, and $\it{x}$=0.35, respectively, measured at 5 K (see Figs.~S4 and S5 in the Supplemental Material~\cite{Supple} for the isotherms at other temperatures). From the inset of Fig.~\ref{Fig3}(a) we observe significant hysteresis for  Mn$_3$Sn at 300 K, similar to a previous report where a weak in-plane ($H\perp z$) ferromagnetism is observed with an effective magnetization of $0.004$ $\mu_{B}/f.u.$~\citep{nakatsuji2015large}. Origin of this weak ferromagnetism is understood from the kagome lattice distortion or off-stoichiometry of the composition~\citep{nakatsuji2015large,sung2018magnetic}. On the other hand, the out-of-plane magnetization ($H\parallel z$) changes linearly with applied field like a typical AFM system.
We further notice that substituting Fe does not significantly affect the out-of-plane magnetic structure at room temperature, however the in-plane weak ferromagnetism increases with Fe concentration as demonstrated in the insets of Figs.~\ref{Fig3} (a), (c) and (e), where one can see increase in spontaneous magnetization from 4$\times$ 10$^{-3}$ $\mu_B/f.u.$ for $\it{x}$=0, 8$\times$ 10$^{-3}$ $\mu_B/f.u.$ for $\it{x}$=0.25  to 1$\times$ 10$^{-2}$ $\mu_B/f.u.$ for $\it{x}$=0.35. Here, doping with Fe mainly increases the in-plane lattice distortion as Fe and Mn have different atomic sizes, and there by enhancing the spontaneous magnetization~\cite{PhysRevLett.100.167203,APOSTOLOV2019113692}.

Next examining the isotherms measured at 5 K [see Figs.~\ref{Fig3}(b), (d), and (f))], from Mn$_3$Sn, the M(H) curves look almost linear for both $H\parallel z$ and $H\perp z$ except that a small hysteresis due to spin-glass transition is observed. On the other hand, from the $\it{x}$=0.25 and $\it{x}$=0.35 systems we observe significant changes in the out-of-plane magnetic structure with Fe doping as the coercivity and spontaneous magnetization increase dramatically, while changes in the in-plane magnetic structure are minimal with Fe doping though the M(H) loop modifies from linear to sigmoid-like ingoing from $\it{x}$=0 to $\it{x}$=0.35. For $\it{x}$=0.35, the out-of-plane magnetization (1.8 $\mu_B/f.u.$) and coercivity (4 T) are extremely large compared to the in-plane magnetization (1 $\mu_B/f.u.$) and coercivity (50 Oe), demonstrating an enhanced magnetic anisotropy with Fe doping. From a closer observation on the M(H) curve of $\it{x}$=0.25, we find field induced asymmetric M(H) loop  [see inset of Fig.~\ref{Fig3}(d)], also observed previously on the metamagnetic Heusler alloy Ni$_{50}$Mn$_{35}$In$_{15-x}$B$_{x}$~\cite{PhysRevB.90.064412} and spin-glass SrRuO$_3$/SrIrO$_3$ superlattice~\cite{doi:10.1021/acsami.7b00150}, possibly due to metastable magnetic phase churned out by Fe doping.

\begin{figure}
\includegraphics[width=\linewidth]{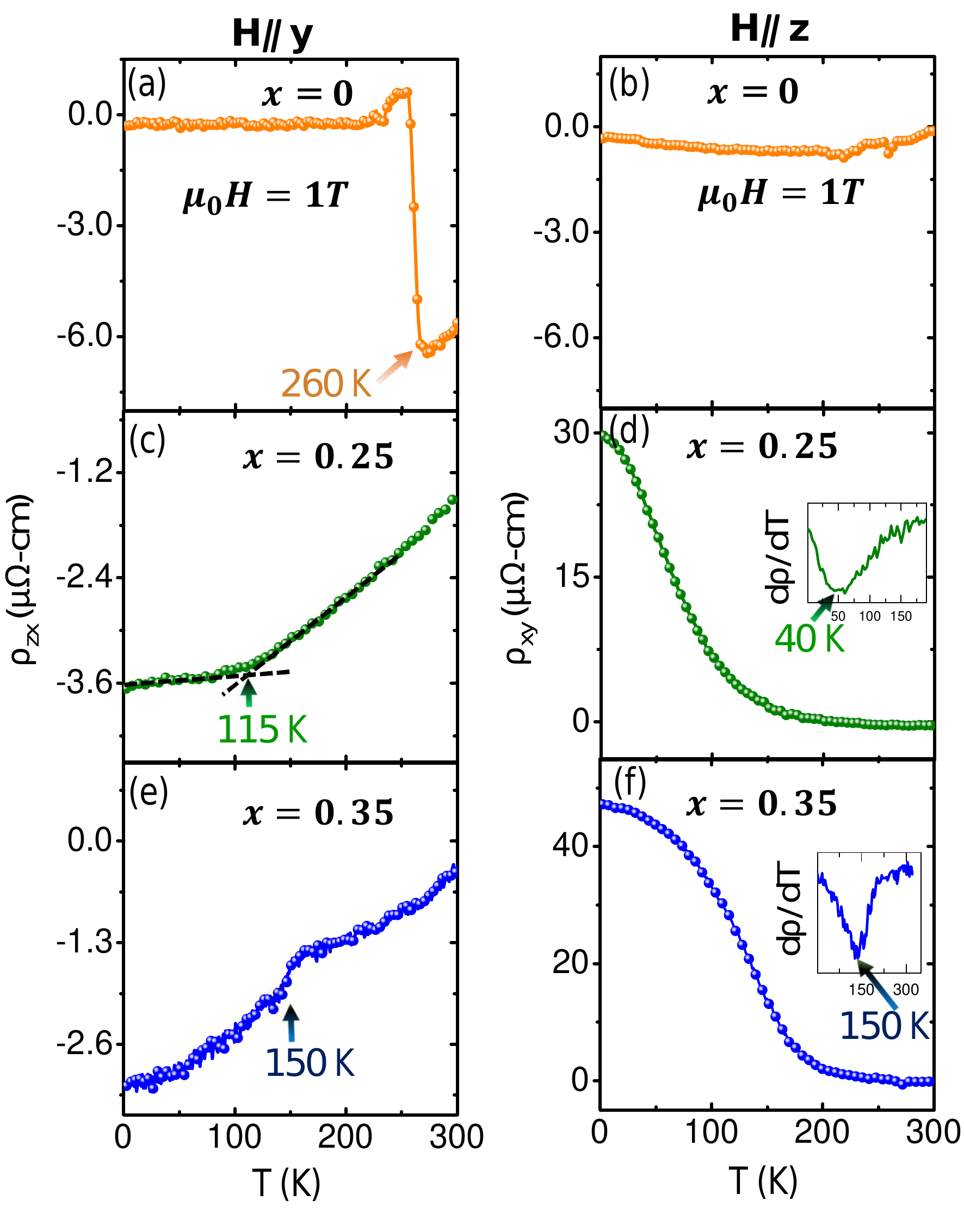}
\caption{\label{Fig5} Out-of plane Hall resistivity ($\rho_{zx}$) plotted as a function of temperature measured under the field of 1T  from (a) $\it{x}$=0, (c) $\it{x}$=0.25, and (e) $\it{x}$=0.35. (b), (d), and (f) show in-plane Hall resistivity ($\rho_{xy}$) plotted as a function of temperature measured under the field of 1T  from $\it{x}$=0, $\it{x}$=0.25, and $\it{x}$=0.35, respectively.  Insets of (d) and (f) show the first derivative of their Hall resistivity with respect to temperature. }
\end{figure}

To quantify the magnetic anisotropy induced by the Fe doping,  we plotted the coercivity measured at various sample temperatures for both $H\parallel z$ and $H\perp z$ orientations in Fig.~\ref{Fig4}. As can be seen from the top panel of Fig.~\ref{Fig4}, for $\it{x}$=0,  there is not difference between the in-plane and out-of-plane coercivities  and almost constant  from 300 K down to 50 K. However below 50 K, though isotropic, it increases drastically with decreasing temperature. For $\it{x}$=0.25, again we find almost constant and nearly zero in-plane and out-of-plane coercivities from 300 K down to 50 K.  Below 50 K,  both in-plane and out-of-plane coercivities increase with decreasing temperature. However, the out-of-plane coercivity ($H\parallel z$ ) is almost three times higher than the in-plane coercivity ($H\perp z$ ) at 2 K. Similarly, in the case of $\it{x}$=0.35,  the in-plane coercivity always found to be zero for all the measured temperatures, while the out-of-plane coercivity starts increasing from zero at 100 K to 4 T at 2 K.  Further, we estimate the magnetocrystalline anisotropy energy density ($K_U$) using the Stoner-Wohlfarth model~\citep{stoner1948mechanism,coey2010magnetism} by considering the $z$-axis as the easy axis (uniaxial) of magnetization. Concentrating only at the low temperature (2 K) where the coercivity is maximum, uniaxial magnetocrystalline anisotropic energy density (K$_U$) is calculated to be 1.2$\times~10^4~J/m^3$ for $\it{x}$=0.25 and 5.3 $\times~10^5~J/m^3$ for $\it{x}$=0.35. These values are significantly higher compared to previously reported in-plane magnetocrystalline anisotropic energy density of $\sim 10^3\: J/m^3$ from Mn$_3$Sn~\citep{duan2015magnetic}.

\begin{figure}
\includegraphics[width=\linewidth]{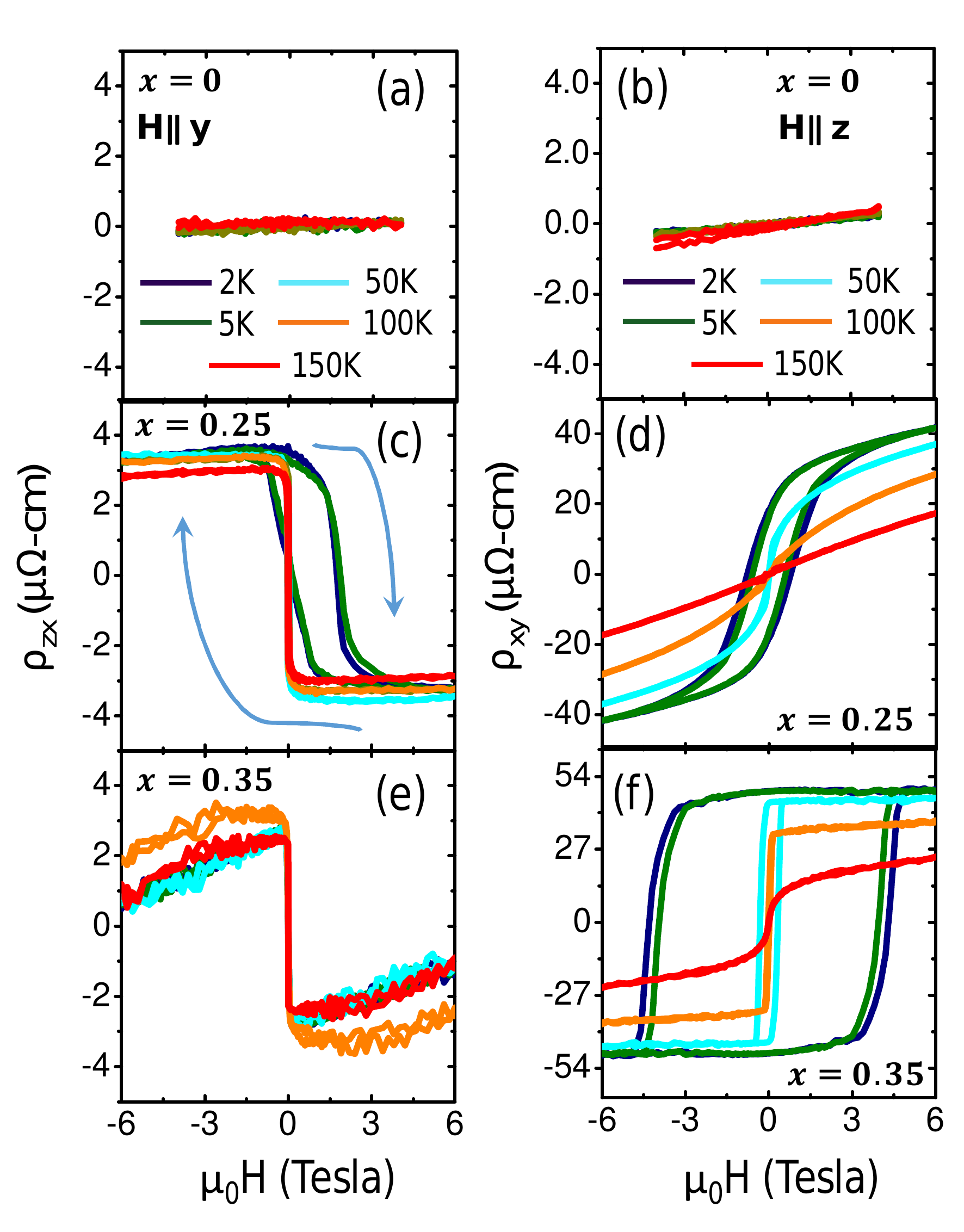}
\caption{\label{Fig6} Field dependent out-of-plane Hall resistivity $\rho_{zx}$ measured at different temperatures from $\it{x}$=0 (a), $\it{x}$=0.25 (c), and $\it{x}$=0.35 (e). Curved blue arrows in (c) indicate field sweeping direction. Similarly, field dependent in-plane Hall resistivity $\rho_{xy}$ measured at different temperatures from $\it{x}$=0 (b), $\it{x}$=0.25 (d), and $\it{x}$=0.35 (f).}
\end{figure}

The effect of Fe doping on the topological properties of Mn$_3$Sn is studied by performing Hall measurements at low temperatures. As shown in Fig.~\ref{Fig5},  we measured out-of-plane [$\rho_{zx}(T)$] and in-plane [$\rho_{xy}(T)$] Hall resistivities  as a function of temperature at an applied magnetic filed of 1 T. From Fig.~\ref{Fig5}(a) we can notice that $\rho_{zx}$ of Mn$_3$Sn increases (negatively)  with decreasing temperature from 300 K down to T$_{SR}$=260 K, and below $\rho_{zx}$ suddenly becomes zero. On the other hand, from Fig.~\ref{Fig5}(b) we can see that $\rho_{xy}$ of Mn$_3$Sn is almost constant from 300 K down to 2 K. Fig.~\ref{Fig5}(c) depicts $\rho_{zx}$ of $\it{x}$=0.25 in which we can notice an increase (negatively) in Hall resistivity linearly with decreasing temperature down to T$_{SR}$=115 K, and below this temperature $\rho_{zx}$ slightly increases to -3.6 $\mu\Omega$-cm at 2 K. Fig.~\ref{Fig5}(d) depicts $\rho_{xy}$ of $\it{x}$=0.25 in which we can notice slight increase (positively) in Hall resistivity with decreasing temperature and reaches to 30 $\mu\Omega$-cm at 2 K. Inset in Fig.~\ref{Fig5}(d) represent $d\rho_{xy}/dT$ having a minima at 40 K, consistent with the ferromagnetic transition temperature of T$_{C^*}$=40 K [see Fig.~\ref{Fig2}(d)]. Fig.~\ref{Fig5}(e) depicts $\rho_{zx}$ plotted for $\it{x}$=0.35 in which we can notice increase (negatively) in Hall resistivity linearly with decreasing temperature down to 2 K, except that a hump at T$_{C}$=150 K is observed.  Fig.~\ref{Fig5}(f) depicts $\rho_{xy}$ of $\it{x}$=0.35 in which we notice increase (positively) in Hall resistivity with decreasing temperature and gets saturated to 48 $\mu\Omega$-cm at 2 K. Inset in Fig.~\ref{Fig5}(f) depicts $d\rho_{xy}/dT$ showing minima at the ferromagnetic transition temperature of T$_C$=150 K [see Fig.~\ref{Fig2}(f)].

From the Hall measurements shown in Fig.~\ref{Fig5}(a) it is clear that the large Hall resistivity ($\rho_{zx}$) observed from Mn$_3$Sn at high temperatures is mainly driven by the nonzero $k$-space Berry phase~\citep{nakatsuji2015large,yang2017topological}. However, at low temperatures ($<$ 260 K), the Berry phase disappears due to spin reorientation, and thus, $\rho_{zx}$ is negligible. Whereas the Berry phase is always absent in the $xy$ plane of Mn$_3$Sn, $\rho_{xy}$ is in general negligible~\citep{nakatsuji2015large}.  In this study, we notice T$_{SR}$=115 K for $\it{x}$=0.25 while no spin-reorientation transition is observed for $\it{x}$=0.35. Worth to mention here that the authors have found spin-reorientation transition at 125 K in the case of Mn$_{2.8}$Fe$_{0.2}$Sn~\cite{Low2021}. Thus, linear increase (negative) in $\rho_{zx}$ with decreasing temperature down to 115 K for $\it{x}$=0.25 and down to 2 K for $\it{x}$=0.35 is due to nonzero $k$-space Berry phase [see Figs.~\ref{Fig5}(c) and ~\ref{Fig5}(e)].    Astonishingly, the in-plane Hall resistivity ($\rho_{xy}$) has been gigantically enhanced with Fe doping as high as 30 $\mu\Omega$-cm for $\it{x}$=0.25 and 48 $\mu\Omega$-cm for $\it{x}$=0.35 at 2 K. Note here that the in-plane Hall resistivity ($\rho_{xy}$) is due to the ferromagnetism induced by the Fe doping.


\begin{figure*}
\includegraphics[width=0.9\linewidth]{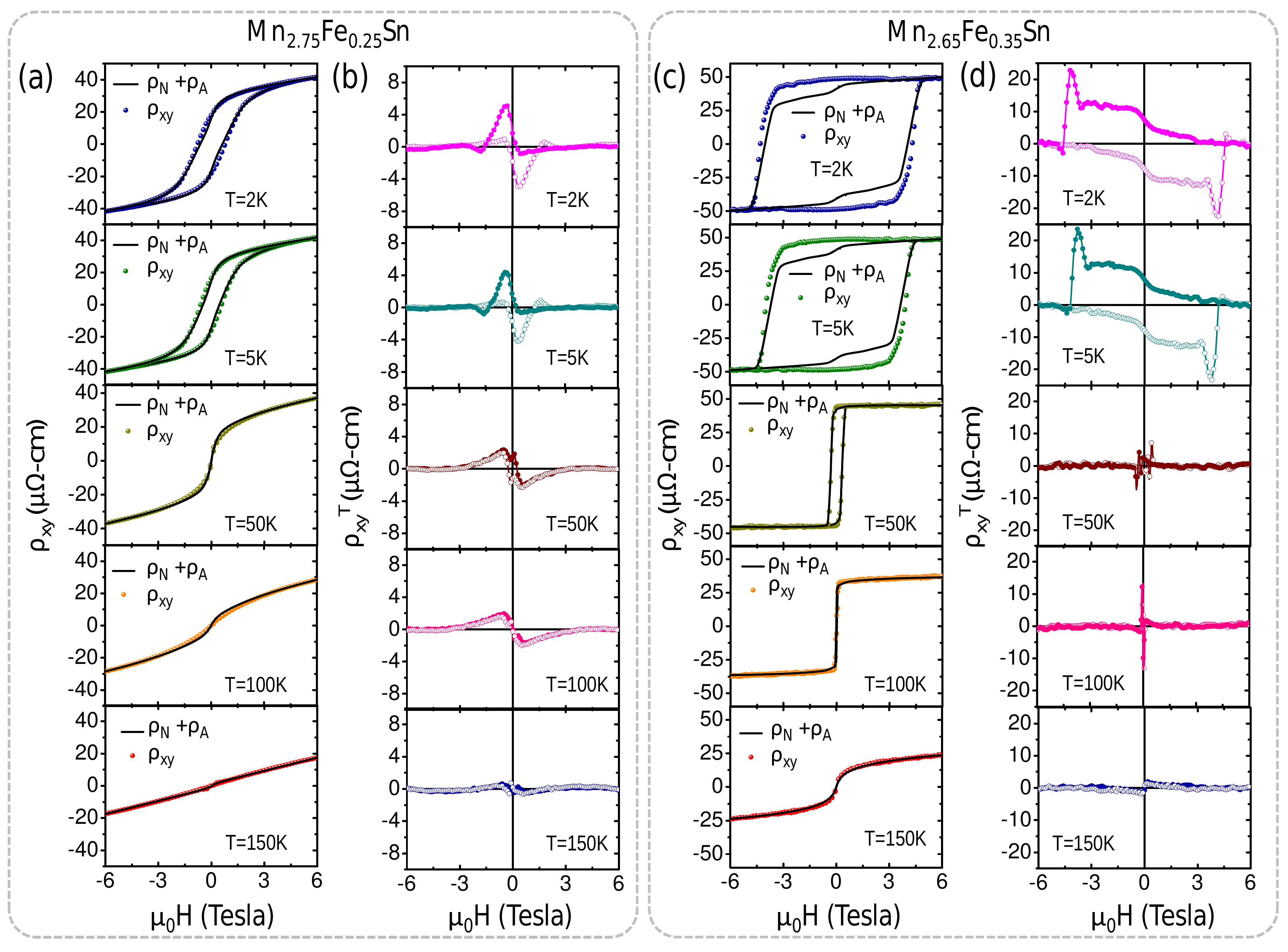}
\caption{\label{Fig7}  (a) and (c) are the field dependent in-plane Hall resistivity ($\rho_{xy}$)  measured at various temperatures from $\it{x}$=0.25 and $\it{x}$=0.35, respectively. The circles are experimental data and the solid black curves are fits to the equation $\rho_{H}=\rho_{N}+\rho_{A}=\textrm{R}_0\mu_{0}\textrm{H}+\textrm{R}_S\textrm{M}$ (see the text for more details). (b) and (d) show the derived in-plane topological Hall resistivity ($\rho^{T}_{xy}$) from $\it{x}$=0.25 and $\it{x}$=0.35, respectively. See Fig.S7 in the Supplemental Material~\cite{Supple} for the raw Hall and magnetoresistance (MR) data.}
\end{figure*}

Field dependent Hall resistivity, $\rho_{zx}(H)$ and $\rho_{xy}(H)$,  from all the samples measured at different temperatures are shown in Fig.~\ref{Fig6}. Here we mainly focus at low temperatures ($\leq$ 150 K) as the doping induced ferromagnetism significantly increases the Hall resistivity at low temperatures. Since below 260 K,  Mn$_3$Sn is already in the spin-spiral texture we find only the ordinary Hall effect as the out-of-plane Hall resistivity ($\rho_{zx}$) linearly depends on the field [see Fig.~\ref{Fig6}(a)]. Similarly, we again find ordinary Hall effect from the in-plane Hall resistivity ($\rho_{xy}$) as well [see Fig.~\ref{Fig6}(b)]. Fig.~\ref{Fig6}(c) depicts $\rho_{zx}(H)$ of $\it{x}$=0.25 in which we observe anomalous Hall resistivity (AHR) with a sudden jump near zero field.   Consistent with asymmetric M(H) data of $\it{x}$=0.25 shown in Fig.~\ref{Fig3}(d), asymmetric hysteresis in AHR is noticed at low temperatures (2 and 5 K).  As for $\rho_{xy}(H)$ of $\it{x}$=0.25 shown in Fig.~\ref{Fig6}(d), we observe sign reversed anomalous Hall resistivity compared to $\rho_{zx}(H)$. Moreover, low temperature $\rho_{xy}(H)$ curves show hysteresis in AHR resembling  the M(H) data of a typical ferromagnetic system  [see $H\parallel z$ data in Fig.~\ref{Fig3}(d)]. Fig.~\ref{Fig6}(e) depicts $\rho_{zx}(H)$ of $\it{x}$=0.35 in which we observe an anomalous Hall resistivity with a sudden jump near zero field, similar to $\it{x}$=0.25. However, unlike in $\it{x}$=0.25, we do not find asymmetric hysteresis in $\rho_{zx}(H)$ at low temperatures. Although at zero field we observe a sudden jump in $\rho_{zx}$,  at higher fields $\rho_{zx}$ is dominated by the ordinary Hall effect as it linearly depends on the field. Fig.~\ref{Fig6}(f) depicts $\rho_{xy}(H)$ of $\it{x}$=0.35 in which we observe sign reversed anomalous Hall effect compared to $\rho_{zx}(H)$, again similar to the case of $\it{x}$=0.25. Moreover, the $\rho_{xy}(H)$ curves show hysteresis in AHR resembling the M(H)data of a typical ferromagnetic system [see $H\parallel z$ data in Fig.~\ref{Fig3}(f)]. From Figs.~\ref{Fig6}(d) and ~\ref{Fig6}(f) we observe increase in AHR hysteresis with decreasing temperature in both $\it{x}$=0.25 and $\it{x}$=0.35 systems.

The Hall resistivity of a ferromagnetic metal can be expressed as $\rho (H)=\rho^{N}(H)+\rho^{A}(H)=\mu_0R_0H+R_sM$~\cite{PhysRev.95.1154}, where $R_0$ is normal Hall coefficient which is inversely proportional to carrier density ($R_0=1/nq$), $R_s$ is anomalous Hall coefficient, and $M$ is  magnetization. However, we are unable to fit properly $\rho_{xy}(H)$ of $\it{x}$=0.25 and $\it{x}$=0.35 using this formula at low temperatures as demonstrated in Figs.~\ref{Fig7}(a) and ~\ref{Fig7}(c). This is because $\rho_{xy}(H)$ has significant contribution from the topological Hall resistivity $\rho^{T}_{xy}$ which can extracted using the formula $\rho^{T}_{xy}(H)=\rho_{xy}(H)-\rho^N (H)-\rho^A (H)$ as plotted in Figs.~\ref{Fig7}(b) and ~\ref{Fig7}(d) for $\it{x}$=0.25 and $\it{x}$=0.35, respectively.
A maximum $\rho^{T}_{xy}$ of 5 $\mu\Omega-cm$ has been derived at a critical field of 0.35 T from $\it{x}$=0.25 and 22 $\mu\Omega-cm$ has been at a critical field of 4 T from $\it{x}$=0.35 at 2 K. The maxima of $\rho^{T}_{xy}$ decreases with increasing temperature and disappears at around 150 K in $\it{x}$=0.25 [see Fig.~\ref{Fig7}(b)], while the maxima of  $\rho^{T}_{xy}$ in $\it{x}$=0.35 becomes negligible above 50 K [see Fig.~\ref{Fig7}(d)].


The topological Hall resistivity is generated by the itinerant electrons by acquiring real space Berry curvature when passing through the nontrivial spin structure of a scalar spin chirality $\chi_{ijk}=(\delta \bm{S_i}\: .[\delta \bm{S_j} \times \delta \bm{S_k} ])$~\citep{PhysRevLett.102.186602,denisov2018general}. The nontrivial spin structure is topologically protected and produces skyrmion lattice, characterised by the nonzero topological charge called the winding number Q~\cite{dupe2014tailoring,Spontaneous_atomic}. By taking into account the skyrmionic picture, we can estimate the skyrmion density ($n_{sk}$) using the relation $B_\textrm{eff}=\phi_0 n_\textrm{sk}$~\citep{PhysRevLett.106.156603,denisov2018general,zeissler2018discrete}. Here, $\phi_0$ is magnetic flux quantum ($h/e$) and $B_\textrm{eff}$ is the effective magnetic field. Also, the topological Hall resistivity $\rho^{T}$ is directly related to $B_\textrm{eff}$ as $\rho^{T}\approx PR_0B_\textrm{eff}$~\cite{PhysRevLett.102.186602,PhysRevLett.106.156603,PhysRevB.91.041122,denisov2018general}. Here,  $P$ is the local spin polarization of charge carrier ($P=\mu_\textrm{spon}/\mu_\textrm{sat}$)~\citep{PhysRevLett.102.186602,denisov2018general}. From the fittings shown in Fig.~\ref{Fig7}(a) we derived $R_0$=$-1\times10^{-9}$  m$^3$/C and from the M(H) data shown in Fig.~\ref{Fig3}(d) we estimated $P=0.07$ for $\it{x}$=0.25. Here,$\mu_\textrm{spon}$=0.659 $\mu_B/f.u.$ taken from Fig.~\ref{Fig3}(d) and $\mu_\textrm{sat}$=9 $\mu_B/f.u.$ (3 $\mu_B/Mn$)~\cite{Brown1990, nakatsuji2015large}. Using these values, $B_\textrm{eff}$ is calculated to be 720 T. Also, the skyrmion density is estimated to $n_\textrm{sk}$ $\approx$ 1.7$\times 10^{17}$ m$^{-2}$ with a skyrmion size (helical period) of about $\lambda_{sk}\approx 2.4\ nm$. Similarly, for Mn$_{2.65}$Fe$_{0.35}$Sn, by taking the fitted value of $R_0=-1.5\times 10^{-9}$ m$^3/$C and calculated value of $P=0.19$, the estimated effective field $B_\textrm{eff}$ $\approx$ 770 T which gives a skyrmion density $n_\textrm{sk}$ $\approx 1.9\times 10^{17}$ m$^{-2}$ and the skyrmion size $\lambda_{sk}\approx 2.2\ nm$. The skyrmion size is estimated using the formula $\lambda_{sk}=(\frac{h}{e}~\frac{\sqrt{3}}{2B_{eff}})^{\frac{1}{2}}$~\cite{kurumaji2019skyrmion}. For $\it{x}$=0.35, we used $\mu_\textrm{spon}$=1.76 $\mu_B/f.u.$ taken from Fig.~\ref{Fig3}(f) and $\mu_\textrm{sat}$=9 $\mu_B/f.u.$ (3 $\mu_B/Mn$)~\cite{Brown1990, nakatsuji2015large}. Note here that we presumably considered the same atomic magnetic moment of 3 $\mu_B/atom$ for both Mn and Fe~\cite{Sales2014}. Further, by assuming that $\approx$ 11.4\% ($\it{x}$=0.35) of Fe doing into Mn$_3$Sn does not significantly change the lattice parameters~\cite{Recour2008} we estimated the skyrmion density of $9\times 10^{17}$ m$^{-2}$ for $\it{x}$=0.25 and  $12\times 10^{17}$ m$^{-2}$ for $\it{x}$=0.35. skyrmion density values estimated from the crystal structure information are in good agreement with the values estimated from the topological Hall resistivity data.

The skyrmion sizes ($\lambda_{sk}$),  2.4 nm from $\it{x}$=0.25 and 2.2 nm from $\it{x}$=0.35,  obtained in this study are comparable to the skyrmion size of 2.49 nm observed from Gd$_2$PdSi$_3$~\cite{kurumaji2019skyrmion}. On the other hand, the helical period of $\approx$ 1 nm is obtained from polycrystalline Mn$_3$Sn~\cite{PhysRevB.99.094430} which is a factor of 2.4 smaller compared to the size of Mn$_{2.75}$Fe$_{0.25}$Sn. However, the B20 systems in their bulk form show much longer-period helical structures (skyrmions). For instance, MnSi shows a helical period of $\approx$18$\ nm$~\cite{Ishikawa1976},  FeGe shows a helical period of $\approx$70 $\ nm$~\cite{Lebech1989}, and Cu$_2$OSeO$_3$ shows a helical period of $\approx$62 $\ nm$~\cite{Adams2012}. Worth to mention here that all three above B20 systems show very large helical periods compared to our studied samples of Mn$_{2.75}$Fe$_{0.25}$Sn and Mn$_{2.65}$Fe$_{0.35}$Sn.

Several mechanisms are proposed to understand the stabilization of the skyrmion lattice in solids such as (i) DM interaction in noncentrosymmetric systems~\cite{PhysRevB.91.041122, park2018magnetic, PhysRevLett.119.087202}, (ii) chiral domain wall induced skyrmion lattice~\cite{Gudnason2014, Cheng2019, Nagase2021, Yang2021}, and (iii)  uniaxial magnetocrystalline anisotropy in the centrosymmetric systems~\cite{yu2014biskyrmion, hou2017observation,doi:10.1073/pnas.1118496109, Preissinger2021}.  It is quite possible to tune the magnetic anisotropy by doping with alien atoms~\cite{doi:10.1021/jacsau.1c00142}. For instance, doping Ni or Pd  in Fe$_3$P at the Fe site converts the in-plane anisotropy to the uniaxial out-of-plane anisotropy and thus the formation of antiskyrmions~\cite{https://doi.org/10.1002/adma.202108770} and Li ion doped in Mott insulator  La$_2$CuO$_4$ acts as a vortex center for the skyrmion~\cite{PhysRevLett.106.227206}. Overall, our study demonstrates that Fe doping into Mn$_3$Sn converts the in-plane magnetic structure of Mn$_3$Sn to a uniaxial anisotropic magnetic structure having easy magnetic axis in the $z$-axis. Hence, Fe doping generates sufficient uniaxial magnetocrystalline anisotropy in Mn$_3$Sn in addition to the geometrical frustration which help to stabilize the low temperature skyrmion lattice.

\section{Summary}

In summary, we have thoroughly examined the effect of Fe doping on the electrical resistivity, magnetic, and topological properties of Mn$_3$Sn. Low temperature magnetic structure has been significantly modified with Fe doping. Especially, we find that Fe doping induces ferromagnetism below 240 K in $\it{x}$=0.25 and below 150 K in $\it{x}$=0.35. In addition, we observe ferromagnetism driven metal-insulator transition in Fe doped systems that is absent in Mn$_3$Sn. We further observe $e-e$ scattering driven resistivity upturn at low temperature with a resistivity minima at 50 K in $\it{x}$=0.35. Fe doping induces magnetic anisotropy such way that the easy magnetisation axis shifts from in-plane in Mn$_3$Sn to the out-of-plane ($z$-axis) in $\it{x}$=0.35. Thus, the large uniaxial magnetocrystalline anisotropy in addition to the competing magnetic interactions at low temperature produces nontrivial spin texture which is responsible for the induced low temperature THE with Fe doping. Further, the topological properties of Mn$_3$Sn are very sensitive to the Fe doping. That means, although Mn$_3$Sn does not show THE at low temperatures, 8\% ($\it{x}$=0.25) of Fe doping shows topological Hall resistivity as high as 5$\mu\Omega-cm$  at a critical field of 0.35T and 11.4\% ($\it{x}$=0.35) of Fe doping shows topological Hall resistivity as high as 22$\mu\Omega-cm$  at a critical field of 4T when measured at 2 K.

\begin{acknowledgments}
Authors thank Science and Engineering Research Board (SERB), Department of Science and Technology (DST), India  for the financial support [Grant no. SRG/2020/000393].
\end{acknowledgments}

\nocite{*}

\bibliography{Fe_Mn3Sn_ref}

\end{document}